# City-scale dark fiber DAS measurements of infrastructure use during the COVID-19 pandemic

Nathaniel J. Lindsey,[1]* Siyuan Yuan,[1] Ariel Lellouch,[1] Lucia Gualtieri,[1] Thomas Lecocq,[2] Biondo Biondi,[1]

[1]Stanford University, Geophysics Department, Stanford, California, USA

[2]Royal Observatory of Belgium, Seismology and Gravimetry Department, Brussels, Belgium

*Corresponding Author: nlindsey@stanford.edu

**Abstract**

Throughout the recent COVID-19 pandemic when government officials around the world ordered citizens to quarantine inside their homes, real-time measurements about the use of roads, hospitals, grocery stores, and other public infrastructure became vital to accurately forecast viral infection rates and inform future government decisions. Although mobile phone locations provide some information about community-level activity, dense distributed geophysical sensing of ground motions across a city are more complete and also natively anonymous. In this paper, we demonstrate how fiber-optic Distributed Acoustic Sensing (DAS) connected to a telecommunication cable beneath Palo Alto, CA captured seismic and geodetic signals produced by vehicles during the COVID-19 pandemic outbreak and subsequent quarantine. We utilize DAS strain measurements of roadbed deformation caused by local cars and trucks in an automatic template matching detection algorithm to count the number of vehicles traveling per day over a two-month period around the timing of the San Francisco Bay Area shelter-in-place order. Using a segment of the optical fiber near a major grocery store on Sand Hill Road we find a 50% decrease in vehicle count immediately following the order, but data from near Stanford Hospital showed a far more subtle change due to on-going hospital activities. We compare the information derived from DAS measurements to other quarantine response metrics and find a strong correlation with the relative changes reported by Google and Apple using mobile phone data.

**Introduction**

During the recent global coronavirus disease pandemic that began at the end of 2019 (COVID-19), human movement in public spaces became a public health issue as government officials advised "social distancing" and other behavioral interventions to mitigate the spread of the novel coronavirus (Ferguson et al., 2020; Tian et al., 2020). In California, nearly 40 million individuals were ordered to stay indoors by state officials beginning 19-Mar-2020, except as needed to maintain critical infrastructure operations, conduct essential services, and obtain food or personal exercise. In the San Francisco Bay Area, the order came from city and county officials three days earlier on 16-Mar-2020.

During public health emergencies of this magnitude, measuring human activity and how public spaces are being used becomes critical for many reasons throughout all phase of the pandemic. Monitoring activity levels during early stages of a pandemic can provide scope of contact tracing and containment efforts. During quarantine, quantifying public activity feeds-back the required information to public health and government officials to make informed decisions such as if the population is properly following quarantine orders (Wesolowski et al., 2012, Tizzoni et al., 2014, Oliver et al., 2020). Historically, this information has been captured in the months or years following the crisis through in-person surveys, but in the past decade public and private mobile phone data has become a tool to understand where, when and how phases of a particular crisis are unfolding in real time. During the COVID-19 crisis, Apple and Google released Mobility Reports beginning April 4, which aggregated mobile phone location services data from individual users into regional statistics about public infrastructure usage (Apple, 2020; Google, 2020). In the San Francisco Bay Area, these reports signaled that vehicle traffic and in-store purchases dropped by 60 – 80% in the days following the shelter-in-place order, while grocery and pharmacy visits decreased by 20%. Although valuable, the utility of mobile phone data is limited in many different ways. For example, the data analyzed are only collected from the subset of people who have access to large data plans and opt-in to use location services. This bias in data sampling likely biases the reported infrastructure usage statistics according to socioeconomic class, age, and also by region. It also means that the Mobility Reports are relative and incomplete because the statistics are reported as percentage change from background and hence the absolute number of people using a public area is not documented. Second, because mobile phone data are segmented by platform, understanding shelter order impacts in particular sectors of a city, such as a park or a major arterial freeway, requires concatenating multiple, potentially heterogeneous databases. Third, while pseudo-anonymization and aggregation responsibly delivered a public good in this case, users are likely to still be wary of the causal linkage established by monitoring their passively generated mobile phone data.

An alternative, fully anonymous, absolute signal of city-scale human activity is ever-present in the low-level seismic wavefield produced by humans, often called the anthropogenic seismic background. For over two decades, seismologists have referred to these $0.5 - 50$ Hz vibrations as "anthropogenic noise" because they commonly obscure earthquakes and other natural Earth signals (Meremonte et al., 1996). This "noise" is actually comprised of seismic waves excited by a multifarious number of moving sources, including vehicles, trucks and trains, buildings and bridges excited by the wind, and generators, pumps and other motorized systems. Instrumenting an urban area with a few precise inertial seismometers has documented the strong spatial and temporal variabilities associated with the anthropogenic seismic wavefield, from high ground motion amplitudes during the daytime to low ground motion amplitudes during night, weekend, and holiday times; ground motions have even been documented from crowds during sporting matches and between sets of a rock concert (Vidale 2011; Diaz et al., 2017).

Recently, during the COVID-19 pandemic, recordings from over 95 urban seismometers were studied in a worldwide effort to measure how the quarantine affected anthropogenic seismic wavefield (Lecocq et al., 2020; Xiao et al., 2020) with results showing anthropogenic ground motion reductions from 10% to more than 100% relative to pre-quarantine levels measured in average dB around 5 - 40 Hz. Seismic energy generated by anthropogenic sources propagates mostly as surface waves in all directions. Due to the highly dissipative nature of the near surface, these waves also undergo severe anelastic attenuation, which is exponential with frequency. Therefore, conventional seismometers deployed in cities can only produce low-resolution maps of general human activity levels, biased towards sources close to the stations. Densifying the recording array using thousands of seismometers with sensor separations of 100 m can measure general urban infrastructure usage patterns across a city such as greater noise near commute infrastructure for a period of several months (Inbal et al., 2015). However, the cost to deploy and maintain thousands of independent stations each with its own sensor, data-logger and power supply throughout a city is impractical for even a brief experiment, and inconceivable over years.

Distributed Acoustic Sensing (DAS) is an emerging geophysical method that turns optical fibers into dense seismic recording arrays with virtual receiver points spaced every 1 – 10 m along the fiber. A DAS experiment consists of connecting an optoelectronic DAS instrument to one end of a standard telecommunications-grade optical fiber. The DAS instrument sends short laser pulses into the optical fiber and measures the subtle phase shifts of Rayleigh scattered light returning to the detector at a predicted two-way travel time (Posey et al., 2000; Masoudi and Newsom, 2016). In this way, the strain field acting on the fiber coupled to the Earth can be sampled at a meter-scale spatial resolution over tens of linear fiber kilometers. Recently, it was shown that DAS can take advantage of otherwise dark telecommunication fibers (Martin et al., 2017; Jousset et al., 2018; Ajo-Franklin et al., 2019), or even be used in combination with lit fiber networks (Wellbrock et al., 2019), effectively leveraging portions of the used or unused modern utility grid for geophysical sensing. In the dark fiber case, DAS recordings commonly capture 10,000 – 20,000 horizontal fiber-oriented components of the strain field in a time continuous fashion, a scale of seismic acquisition that is not possible with traditional seismometers. An additional advantage for dark fiber DAS is that the instruments are deployed in secure telecommunications utility buildings and the fiber is secure under the road. Therefore, this type of experiment can be rapidly deployed in a few hours and continue to record for years without major operations costs. In urban areas, DAS measurements have already been used for near surface geological imaging (Dou et al., 2016, Martin, 2018, Spica et al., 2019; Fang et al., 2020) and earthquake recording (Lindsey et al., 2017, Martins et al., 2019).

In this paper, we use continuous dark fiber DAS recordings from a telecommunication cable running through Palo Alto, CA, leased from Stanford University IT Services to measure how different public sectors of the city responded to the COVID-19 pandemic quarantine order. We begin by illustrating the diversity of urban soundscapes and the related challenges of disentangling urban ground noise source processes. Then, we show how the broadband nature of DAS buried beside roads in a city captures both high frequency (3 – 30 Hz) anthropogenic surface waves radiated by vehicles as well as a more interpretable geodetic signal of the response of the roadbed to vehicle loading which occurs in a lower frequency range (< 1 Hz). We use the geodetic signals to automatically detect individual vehicles and measure their speed, size, and direction on the road. We compare the number of vehicles passing through portions of the public infrastructure around Palo Alto recorded by DAS to the Google and Apple Mobility Reports and discuss the complementarity of this anonymous geophysical data source.

**Results**

*Exploratory Data Analysis:* Over 14.5 TB of continuous DAS strain data were recorded between 01-Mar-2020 through 01-May-2020 on the Palo Alto DAS array (see map in Figure 1a). The general characteristics of the local urban seismic background were documented for a subset of DAS channels near Stanford Hospital and Sand Hill Road by computing daily power spectral density (PSD) after some basic pre-processing (see Materials and Methods). Most DAS channels were dominated by broad frequency maxima spanning half an octave or more (Figure 1b), but the exact frequency peak varied significantly across the array, even within 100 m. Peak amplitudes from Sand Hill Road were 20 – 30 dB higher than on the non-Sand Hill Road segment. Some channels, like from Upper Sand Hill, also showed minor, narrow frequency peaks. Significant PSD reductions of up to 7 dB were observed at most channels beginning after the first week in March (black traces) to the last day (solid color; transparency decreases with time) in Figure 2b (analyzed in greater detail below). Other channels, including those near Stanford Hospital, showed only one lower frequency peak that did not systematically change during quarantine.

Combining DAS density with multi-kilometer lateral apertures permitted more targeted exploration of the sources responsible for the urban seismic background. For example, Figure 2a documents vehicles driving at ~15 m/s (33.55 mph) on Sand Hill Road, where the speed limit is 15.65 m/s (35 mph). The data in this figure are pre-processed strain time series recorded during a three-minute period from 10-Mar-2020 using the evenly-spaced DAS array, and then bandpass filtered between 3 – 30 Hz. In this high frequency band, ground motion could be tracked coherently 100 – 200 m away from the vehicle, but longer wavelengths likely travel much further. The reduced seismic background at non-Sand Hill Road DAS channels was observed to transition sharply from high to low at X=3.18 km where the fiber turned off of Sand Hill Road towards Stanford Hospital (see Fig. 1a), potentially related to fiber directionality, distance from the road, and coupling.

Filtering the Palo Alto DAS data in a lower frequency range of 0.1 – 1 Hz revealed a localized wavelet pattern with peak amplitudes around 5 – 10 nanostrain, which migrate in time and space at the speed limit for the road (Figure 2b). Jousset et al. (2018) observed a similar low frequency horizontal strain response on optical fibers with a DAS experiment in Iceland and suggested it was related to the quasi-static or geodetic loading of passing vehicles.

In general, vehicular seismic sources can be modeled using a collection of vertical point forces located at the vehicle's wheel-road contacts (Ben Zion & Zhu, 2002; Li et al., 2018, Jousset et al., 2018; Brenguier et al., 2019). The geodetic strain signal from a vehicle is equal to the change in displacement over the DAS gauge length (G). Displacement can be modeled with basic knowledge of the fiber location, vehicle location, vehicle mass, soil shear modulus ($\mu$) and Poisson's ratio ($\nu$) using the Boussinesq approximation for a loading point source:

$$u_x = \frac{F_z}{4\pi\mu}\left(\frac{zx}{r^3} - \frac{(1-2\nu)}{r+z}\left(\frac{y}{r}\right)\right) \quad (1)$$

$$\varepsilon_x = \frac{\Delta u_x}{G} \quad (2)$$

In (1), x and y represent the across-road and along-road distances, respectively, z is the fiber burial depth, $r = \sqrt{x^2 + y^2 + z^2}$, and $F_z$ is the weight of the vehicle. The measured horizontal displacement ($u_x$) is linearly related to the vehicle's mass, and falls off rapidly as 1/r² unlike the higher frequency surface wave ground motions which radiate hundreds of meters beyond the vehicle's location. The point load horizontal strain response ($\varepsilon_x$) described in (2), which is equivalent to the change in (1) divided by the gauge length (i.e., spatial derivative of displacement), thus depends only on the vehicle's speed and Poisson's ratio. However, the strain response to a four-wheeled passenger vehicle, for example, also depends on the inter-wheel separation or vehicle area. Based on a plausible parameter space, we determined that urban vehicles produce geodetic signals with energy around 0.1 - 1 Hz (see Materials and Methods).

***Vehicle Detection and Characterization:*** Vehicle-induced geodetic strains in the range 0.1 – 1 Hz were utilized in a template-matching algorithm to detect and characterize changes in Palo Alto vehicle traffic patterns during the COVID-19 quarantine (Figure 2d, see Materials and Methods for a description of the algorithm). Figure 3 shows the number of vehicles detected per day during the morning commute hours (4:00 am – 12:00 pm) for two selected DAS locations marked in Figure 1. According to our results, the number of vehicles on Sand Hill Road decreased from about 11,000 per day to around 5,500 per day after the 16-Mar-2020 Bay Area shelter-in-place ordered. This is equal to a 40 – 60 % decrease in the measured number of vehicles relative to a background level which we calculate from the median of the first week of the experiment (01-Mar-2020 through 07-Mar-2020, inclusive). Reduced traffic levels persisted through quarantine without increasing. Meanwhile, the number of vehicles detected using DAS data near Stanford Hospital documented a stable level of vehicles around 2,000 vehicles per day in the two-week period after the order. Beginning around 06-Apr-2020, the vehicle activity near Stanford Hospital steadily declined to around 80% of the pre-quarantine level most likely as a result of the closure of the Stanford clinics as well as Stanford Hospital instructing hospital staff to change where they parked their vehicles. Additionally, the dramatic modification of traffic patterns in Palo Alto could have rerouted drivers to a faster route avoiding this part of the city. There were several anomalous times of vehicle activity such as on 17-Apr-2020 (Good Friday) and 25-Apr-2020 when it appears more vehicles were traveling in this area potentially to spend time outdoors on Stanford University's campus or surrounding parks. On 16-Mar-2020, the day of the shelter-in-place order, there is a single-day increase in traffic suggestive of people driving to the store to get supplies.

**Discussion**

One DAS experiment provides highly resolved statistics about public infrastructure utilization across many large sectors of a city through time. According to our results from the Stanford DAS-2 experiment, the COVID-19 quarantine order from local and state government officials resulted in a major decrease in commuter traffic along Sand Hill Road but sustained use of critical infrastructure such as the road near Stanford Hospital. The baseload of vehicle traffic on Sand Hill during the quarantine was likely related to essential business traffic like the major grocery store and pharmacies. These changes may seem obvious as Sand Hill is a commuter corridor connecting the I-280 freeway and CA-82 highway to businesses and workplaces that closed during quarantine while Stanford Hospital remained busy, but accurately quantifying these changes in real-time is a challenge.

A commonly reported metric during the quarantine was derived from mobile phone users. We find a strong correlation between the 50% decrease in vehicles counted using DAS on Sand Hill Road and the 50 - 70% decrease in mobile phone-derived activity data reported by Apple (for Santa Clara County driving data) and by Google (for San Francisco Bay Area retail and recreation data; see Figure 3). The Mobility Report data could be biased downwards from the actual number of vehicles on the street for many reasons. For instance, they exclusively rely on data from a select pool of users, and thus will not include users making trips without data services, public transit vehicles, or police/fire/emergency vehicles, which were all still using Sand Hill Road during quarantine. Furthermore, the Mobility Reports aggregate different regions, which may be less representative of Sand Hill Road.

Over 450,000 automatic vehicle detections were made using DAS on Sand Hill Road during the 8-week experiment. We found that the quasi-static or geodetic strain response to individual vehicle loading/unloading presented clear evidence of a proximate vehicle (Figure 2d). Because this strain falls off like $1/r^2$, vehicles passing far from the fiber such as on the opposite side of the street have a greater likelihood of being missed, as can be seen in Figure 2b where Northbound vehicles are weaker than Southbound ones (fiber is located on the Southbound side). While the template matching algorithm performs better than a simple threshold detection algorithm, more sophisticated methods that utilize machine learning or the distributed nature of DAS data are required to refine these results (Martin et al., 2018).

Spatially-localized roadbed deformations from vehicles passing in the immediate vicinity of the fiber provide a more accurate detail about traffic patterns than is available to point sensor seismic measurements. This is because one urban seismometer, or more commonly a short geophone, only records high frequency anthropogenic surface waves. Such energy could be produced by vehicles traveling anywhere within a radius of several hundred meters. In Figure 4, we show that the DAS data from Sand Hill Road documents a relationship between the number of vehicles and the total surface wave strain energy recorded in two different high frequency bands (4-10 Hz and 15-30 Hz). This relationship provides calibration of the vehicle component of the anthropogenic surface wave background recorded by seismometers.

In times of crisis, access to DAS-derived information about public infrastructure utilization is valuable to public health and government officials because it provides city-wide measurements which are natively anonymized and aggregated with the requisite level of spatial and temporal granularity to make decisions. DAS retrieves data about how infrastructure is being used across a large urban area with an easily deployable instrument.

**Materials and Methods**

***Experimental design:*** The Stanford DAS-2 experiment, which used an OptaSense ODH-3 interrogator connected to a telecommunication fiber running from Stanford University campus to near the Stanford Linear Accelerator (SLAC), began recording on 10-Dec-2019 and at the time of this paper is still recording. The ODH-3 records optical phase shift (proportional to strain) per gauge length. The sampling rate was 250 samples per second. The gauge length was 20 m. There were, in total, 1250 channels with a channel spacing of 8.16 m. About 350 channels were located along the relatively straight Sandhill Road section between the quiet portion of the array near Stanford Hospital (Channel #300) and SLAC (Channel #750). The data volume write rate to-date is approximately 108 GB/day.

***DAS data processing:*** Unless otherwise specified, DAS data processing involved the following procedure: 1) load raw SEG-Y data in multi-hour continuous time series, 2) convert from optical phase to strain based on photonic parameters (amounting to multiplication by 10.77097) yielding data with nanostrain units, 3) remove a linear trend, 4) remove the mean, and then 5) reduce the common mode noise by subtracting a contemporaneous time series recorded by an isolated channel from DAS array. Often, we isolated a particular seismic signal by then bandpass filtering the data. Unless otherwise specific, the bandpass filter applied was a four-corner, zerophase, Butterworth-Bandpass filter. PSD strain values for DAS recordings were computed as $10 log_{10}(\varepsilon_x{}^2)$ and thus have units of nanostrain$^2$/Hz or dB/Hz. RMS strain values for DAS records were obtained by taking the square-root of the integral of the power spectrum computed for a 30-minute window of data according to Parseval's theorem. In Figure 4, we sum 30-minute RMS values to obtain a total strain.

***Model of roadbed deformation due to vehicle loading:*** Small passenger vehicles, light trucks, and larger vehicles (1000 – 3000 kg) traveling at city speeds of 10 – 20 m/s (22 – 45 mph) produce quasi-static/geodetic signals with energy around 0.1 - 1 Hz on roadside fibers that are buried on the order of 1 m below the surface and located at a distance of less than 20 m from where vehicles are located on the road. To compute the modeled synthetic horizontal strain signal shown in Figure 2, we assumed a fiber depth of 2 m, a lateral offset of 10 m, vehicles speeds between 10 and 15 m/s, vehicle mass between 1000 – 3000 kg, and appropriate values for sandy/clayey soils, such as a Poisson's Ratio of 0.4, a Young's modulus of 80 MPa, and a shear modulus of 28.6 MPa.

***Vehicle detection algorithm:*** To determine the number of vehicles driving in different sectors of Palo Alto we analyzed strain recordings from individual DAS channels using a template-matching detection algorithm based on the characteristic geodetic signal described in the main text and modeled in Figure 2d. We used a different template for each channel and day. The 20-second duration template was constructed from a median of 200 SLA/LTA detections. STA/LTA refers to an earthquake detection algorithm in which the ratio of a short-term average (STA) of a time-series over a long-term average (LTA) is employed to detect signal impulses rising above the background noise with a threshold. Classically, the STA/LTA algorithm is computed over the power of the seismic trace. We used a 0.01-s duration STA window and a 10-s duration LTA window, and automatically determined the detection threshold such that about 200 templates were detected per day. These settings were validated against human picks for short segments and it was determined that the STA/LTA algorithm accurately detected the higher amplitude vehicle signals. Using the median of a large population of STA/LTA detections effectively avoided corruption by earthquake signals. We then computed the full cross-correlation between the template and the daily time series, labeling windows with cross-correlation value above 0.7 as vehicles.

**Acknowledgments**

**General:** The authors would like to thank Martin Karrenbach, Victor Yartsev, Lisa LaFlame from Optasense Inc. for loaning the DAS instrument used in this experiment, as well as the Stanford ITS fiber team, and in particular Erich Snow, for crucial help with the Stanford DAS-2 experiment. We also wish to thank the Stanford School of Earth IT team for hosting the interrogator in the Scholl computer room, and Bob Clapp for file system management of the DAS data. We acknowledge the Social Seismology Team for electronic correspondence during the COVID-19 quarantine on ideas related to this project. Plotting and data analysis made use of the Obspy Python package (Krischer et al., 2015). Florence Ho provided information on Stanford Hospital clinic closures. **Funding:** This research was financially supported by affiliates of the Stanford Exploration Project. NL was supported the Thompson Postdoctoral Fellowship; AL was partially supported by the Israeli Ministry of Energy under the program for postdoctoral scholarships in leading universities. **Author contributions:** NL designed the study, conducted all analysis and wrote the manuscript; SY led the Stanford DAS-2 data management and processing and contributed to the manuscript; AL led the Stanford DAS-2 experiment and contributed to the manuscript; LG and BB advised on the study and contributed to the manuscript, TL motivated this study with initial observations and some open source Python codes to analyze ground motion correlations during quarantine. **Competing interests:** There are no competing interests. **Data and materials availability:** DAS data used to determine the results of this paper are available upon request.

**Figures and Tables**

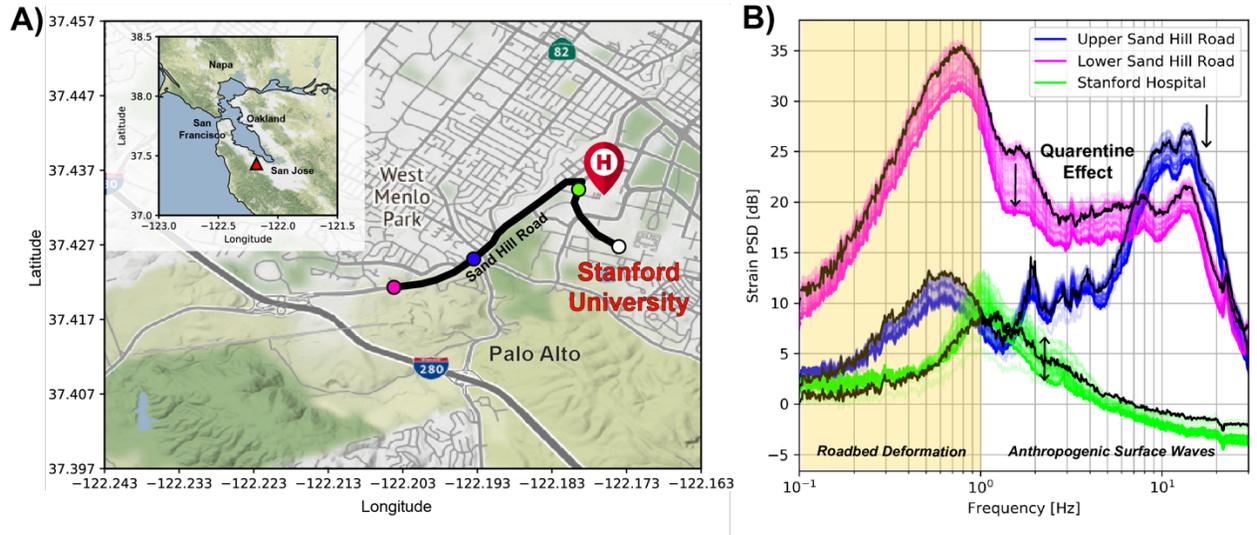

**Fig. 1. Stanford DAS-2 Experiment Array and Anthropogenic Noise Spectra. (A)** Map of optical fiber path used for DAS (black line) located in Palo Alto, California (inset, red triangle). Circles color-code a few DAS channels used for detailed analysis (white = quiet reference; green = near Stanford Hospital; blue = upper Sand Hill Road; magenta = lower Sand Hill Road). **(B)** Daily median power spectral density (PSD) of strain data computed for DAS channels shown in (A) highlight strong spatial variability in anthropogenic ground motion, and how this ground motion changed during quarantine. Black curves represent PSD on 01-Mar-2020. Transparent colored curves fade to solid with increasing experiment date through end of April 2020. Yellow shading highlights frequency band used to identify vehicles with DAS based on their geodetic roadbed deformation.

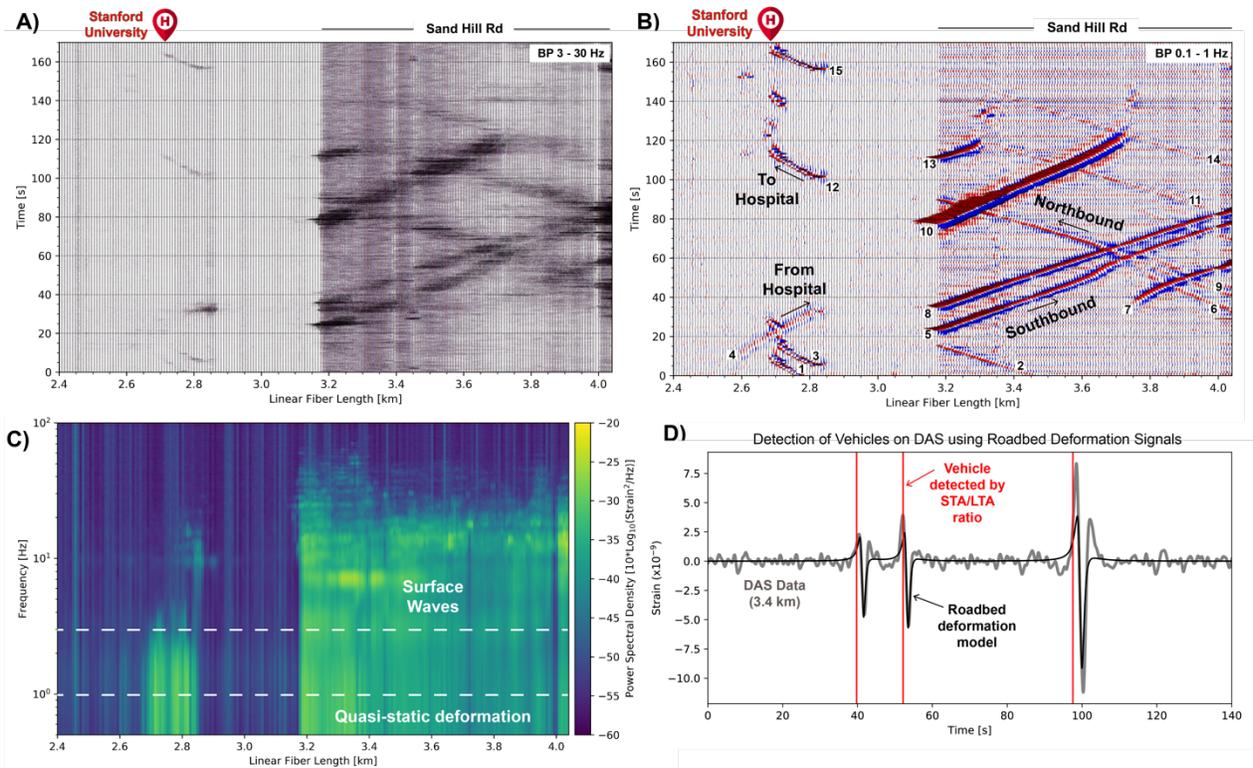

**Fig. 2. Vehicle Observations from Stanford DAS-2 Experiment.** (**A**) Example of DAS recordings bandpassed around 3 – 30 Hz showing vehicle surface waves and contrasting background energy between Stanford Hospital and Sand Hill Road sections. (**B**) Same as (A) but bandpassed around 0.1 – 1 Hz to highlight the high quality geodetic strain responses of the roadbed due to vehicle loading. Individual vehicles are numbered. (C) Continuous wavelet transform applied to spatial axis of unfiltered data shown in (A) and (B) highlighting dominant frequencies of different array segments. (D) Example strain data from DAS channel at 3.4 km low-pass filtered as in (B), plotted with a model of the horizontal strain described in the text (black line), and three STA/LTA detections (red lines) for vehicles #5, #8, #10 shown in (B). A matched template algorithm was then applied using the median of approximately 200 detected vehicle signals and scanning over the full daily time series for the DAS channel.

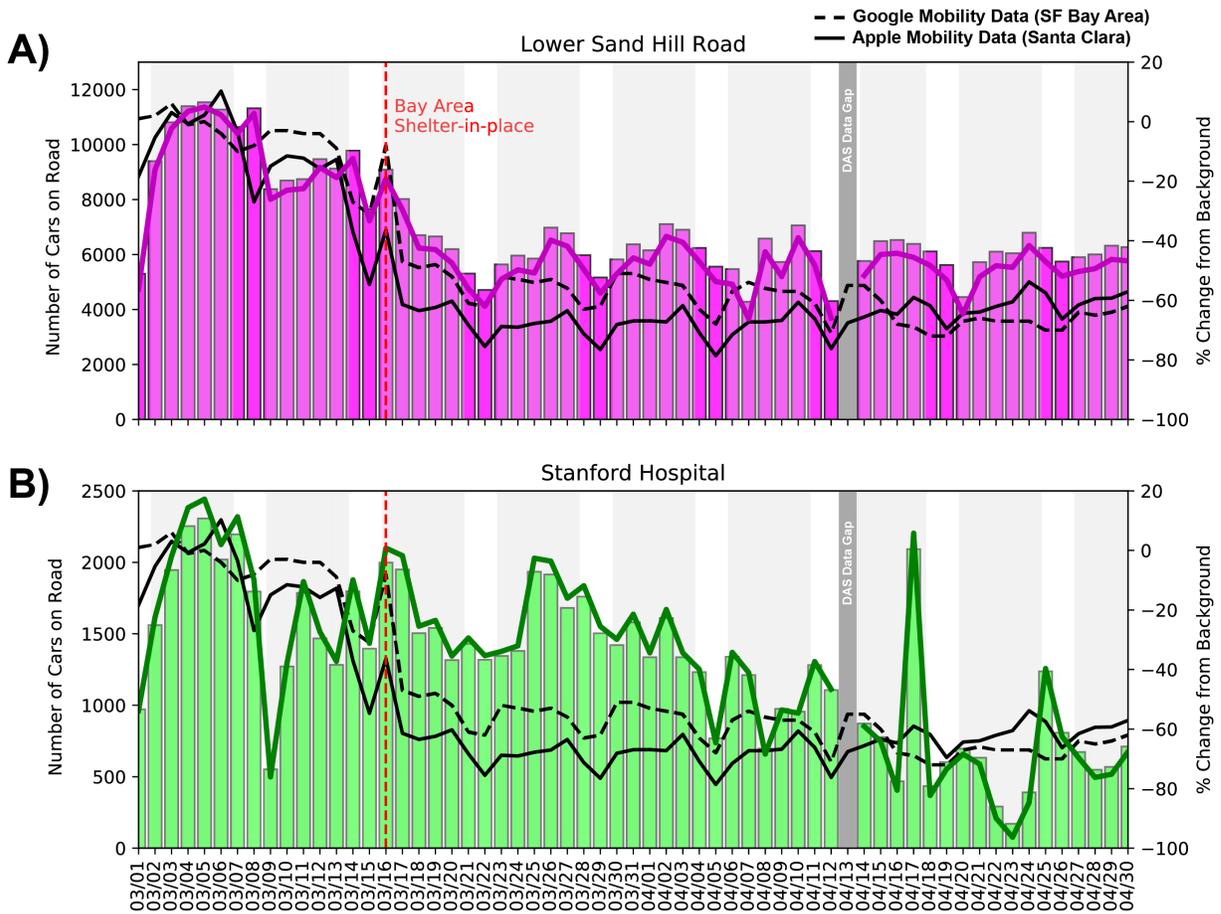

**Fig. 3. DAS vehicle detections measured before and during COVID-19 quarantine.** (A) Number of vehicles (left y-axis) detected near Lower Sand Hill Road during morning commute between 4:00 am and 12:00 pm. Lines show the relative change as a percentage (right y-axis). Magenta line summarizes the change in DAS detections from a 7-day median background during the first week, compared with mobile phone statistics provided by Google (dashed black) and Apple (solid black). Vertical red line indicates timing of Bay Area shelter-in-place. Gray/white background represents weekdays/weekends. One data gap is shown when the DAS instrument was offline for maintenance. (B) Same as (A) for DAS channel near Stanford Hospital.

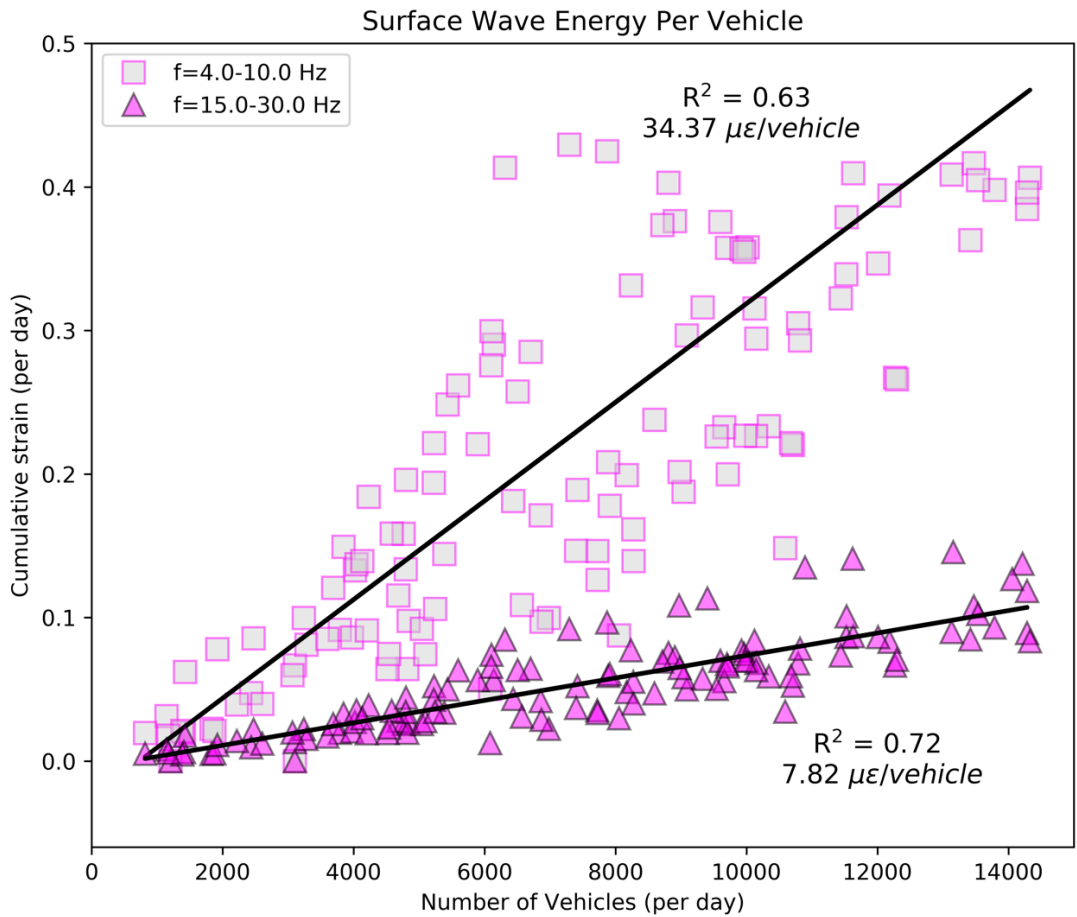

**Fig. 4. Relationship between anthropogenic surface waves and vehicle number.** Total horizontal strain for the daytime hours recorded by a Lower Sand Hill Road DAS channel versus the number of vehicles detected by template matching. Two frequency bands are shown, f=4.0-10.0 Hz (squares) and f=15.0-30.0 Hz (triangles). The energy of surface waves produced by vehicles decays with decreasing wavelength and increasing propagation length, hence the correlation improves at higher frequencies.